\documentclass[a4paper]{article}
\usepackage{graphicx}
\usepackage{epsfig}
\usepackage{float}

\begin{document}

\begin{flushright}
\small  AMS note 2006-03-01\\
\small  LAPP-EXP-2006-01
\end{flushright}
\vspace{30pt}
\begin{center}
\Large Towards a new tool for the indirect detection of Dark Matter : \\building of a SuSy spectrum generator based on \verb"micrOMEGAs" \\
\vspace{30pt} \normalsize Pierre Brun\footnotemark[1]
 \footnotetext[1]{Laboratoire d'Annecy-le-vieux de Physique des Particules, CNRS/IN2P3/Univ. de Savoie, brun@lapp.in2p3.fr}
\end{center}

\vspace{40pt}

\begin{abstract}
\emph{In the quest for indirect signals from dark matter annihilation, powerful computation codes are required. I report here
a new code based on micrOMEGAs devoted to the analysis of such signals in term of Supersymmetry. It computes gamma rays and
positrons fluxes in a general SuSy model, as well as the other charged cosmic rays and neutrinos source terms. This work aims
to propose an alternative to the DarkSUSY code by providing inclusive signals from SuSy for dark matter indirect searches.
Therefore it can be used for sensitivity studies and data analysis.}
\end{abstract}
\vspace{20pt}

\section{Supersymmetric dark matter}

\subsection{Supersymmetry and cosmology : the micrOMEGAs code}

\begin{figure}[H]
\centering
\includegraphics[width=7cm]{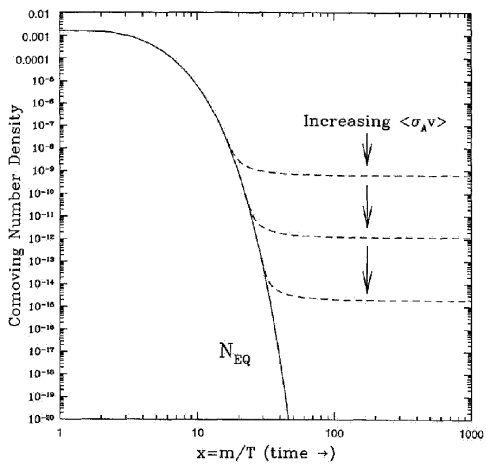}
\caption{Evolution of the co-moving wimp density with temperature}
\label{freeze}
\end{figure}

The Standard Model of particle physics describes two distinct families, bosons and fermions as interaction and matter
particles respectively. The integer and half-integer spins can be merged in a unified model if one assumes that a symmetry
exists between those two sectors, this is the purpose of Supersymmetry (SuSy). In addition to this aesthetic issue, SuSy can
solve some of the Standard Model limitations, such as the stability of the Higgs boson mass as regard to radiative corrections, the
unification of forces or the understanding of the electro-weak symmetry breaking mechanism. Since no SuSy particle has ever
been detected, it must be that the symmetry is broken at today's reached scales, and partners are heavier than the standard
particles. Furthermore, assuming the conservation of a new quantum number called R-parity, SuSy predicts the existence of
relic particles from the Big-Bang, these are \emph{stable, neutral, weakly interacting, massive and non baryonic} (WIMP's).
When it has these characteristics, the Lightest SuSy Particle (LSP) is a good candidate for dark matter. This possibility is
very interesting since WMAP measurement of the cosmic microwave background anisotropies point out that there must be a large
amount of non baryonic dark matter in our Universe. These measurements, in addition to the observation of type Ia supernovae,
large scales structures features and constraints from Big-Bang nucleosynthesis give the following picture of our Universe
content \cite{bennet}:
\begin{itemize}
  \item{
  The universe is globaly flat, with $\Omega_{tot}=1.02\pm0.02$.
  }
  \item{
    The major part of the energy is under the form of a cosmological constant
    $\Omega_{\Lambda}=0.73\pm0.04$.
  }
  \item{
    The matter component is $\Omega_{matter}=0.27\pm0.04$ and is $84\%$ non baryonic
    ($\Omega_{baryons}=0.044\pm0.004$).
  }
\end{itemize}

\verb"micrOMEGAs" purpose is to accurately compute the relic abundance in a general SuSy model \cite{micro} (namely the
Minimal Supersymmetric Standard Model, MSSM). To do so, one has to solve the Botzmann equation:
\begin{equation}
  \frac{dn}{dt}=-3Hn-<\sigma v>(n^{2}-(n^{eq})^{2})
  \label{boltz}
\end{equation}
Where $n$ is the number density of relic particles, $<\sigma v>$ their annihilation cross section and H the Hubble constant.
Basically this equation indicates that the annihilation rate has to be balanced against the expansion of the Universe,
leading to a freeze-out of the LSP. Figure \ref{freeze} shows how this freeze-out occurs, we can see the evolution of the
co-moving wimp density. In the first part, the particles interact sufficiently one with another to make the LSP annihilate,
and the LSP density goes down. At a particular temperature, the Universe expands so that the Hubble radius get greater than
the mean interaction length. From this moment, the annihilation rate is very low and the relic particles density remains
constant.

It happens that some particles have a mass close to the LSP's mass and then contribute to the decrease of the LSP density,
these processes are the so-called $coannihilations$ \cite{pierre_coan}. In that case one has to replace $<\sigma v>$ in eq.
\ref{boltz} by a sum over all coannihilation channels:
\begin{equation}
  \frac{dn}{dt}=-3Hn-\sum_{i,j=1}^{N}<\sigma_{i,j} v_{i,j}>(n_{i}n_{j}-n_{i}^{eq}n_{j}^{eq})
  \label{boltz2}
\end{equation}
One great performance of \verb"micrOMEGAs" is to dynamically include -when they are relevant- all possible coannihilation
channels. The total number of processes that can be involved is about 3000.

\subsection{Supersymmetry breaking}

Once  ``supersymmetrized'',  the  Standard  Model  becomes  the  Minimal  Supersymmetric  Standard  Model (MSSM). if SuSy was
effective, Sparticles would have the same mass as the corresponding particles. Superpartners have never been observed and it
must be that SuSy is a broken symmetry. Then one has to parameterize the breaking by introducing terms in the Lagrangian that
explicitly break SuSy. This breaking imposes the introduction of over 100 free parameters. In order to make physical
predictions, one can fix them or assume a model for SuSy breaking that generates these soft SuSy breaking terms. In this
note, we will work in two SuSy breaking models frameworks :

\begin{itemize}
\item{Minimal SUperGRAvity, in which SuSy is broken via gravitationnal interaction. In the mSUGRA model,
  all physical quantities can be derived from 5 parameters specified
  at the Planck scale, these are :
  \begin{itemize}
  \item{$m_{0}$: common scalar mass}
  \item{$m_{1/2}$:common fermion mass}
  \item{$A_{0}$: universal trilinear couplings}
  \item{$tg(\beta)$: ratio of the neutral Higgs vacuum expected values}
  \item{$sgn(\mu)$: sign of the Higgs mass parameter $\mu$}
  \end{itemize}
}

\item{Amonaly Mediated SuSy Breaking, in which SuSy is broken via the super-Weyl anomaly effects \cite{randall}
  In AMSB, 4 Planck scale parameters are enough to describe the MSSM :
  \begin{itemize}
  \item{$m_{0}$: common scalar mass}
  \item{$m_{3/2}$: gravitino mass}
  \item{$tg(\beta)$: ratio of the neutral Higgs vacuum expected values}
  \item{$sgn(\mu)$: sign of the Higgs mass parameter $\mu$}
  \end{itemize}
}
\end{itemize}

In those frameworks, the LSP is the lighest neutralino $\tilde{\chi}_{1}^{0}$ (or $\chi$), a mixing of gauge bosons partners.
In this note only mSUGRA and AMSB are considered, but the code also allows to work in the Gauge-Mediated SuSy Breaking
scenario.

\subsection{Cosmological constraints on mSUGRA}

\begin{figure}[H]
\centering
  \includegraphics[width=10cm]{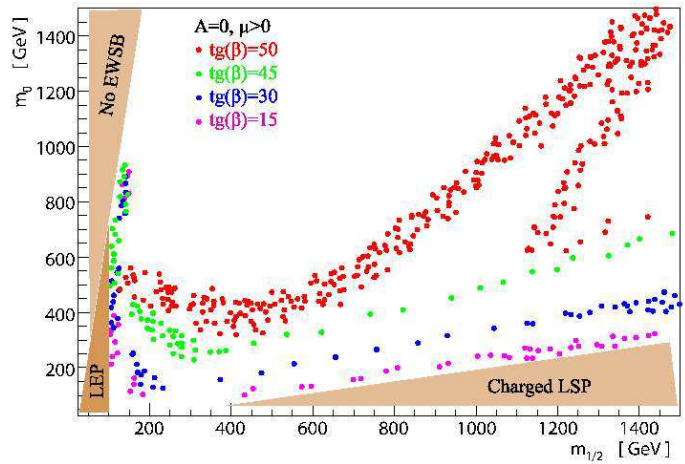}
  \caption{Cosmological constraints on gravity broken SuSy}
  \label{cosmo}
\end{figure}

The initial goal of \verb"micrOMEGAs" is to compute accurately the relic density, and compare it to cosmological measurements
thus putting constraints on the parameter space of a given model. To illustrate this, a scan of the parameter space has been
performed. On Figure \ref{cosmo}, only the WMAP $4\sigma$ allowed points are plotted in the $m_{0}$-$m_{1/2}$ plane, $i.e.$
when the computed thermal relic density satisfies $0.076\leq \Omega h^{2}\leq 0.148$. The mass spectrum generator used here
is \verb"SuSpect" \cite{suspect}.

\section{AMS02 potential for indirect dark matter searches}

AMS02 is a particle physics spectrometer to be placed on the International Space Station for 3 years. The detector allows to
measure charged cosmic rays as well as $\gamma$ ray fluxes in the range of 1 GeV to a few TeV, including particle
identification, charge reconstruction, isotopes separation in case of light nuclei \cite{alcaraz}. It consists in different
specific sub-detectors, with some redundancy in measurements. A Silicon Tracker surrounded by a superconducting magnet
provides charge and rigidity, two planes of a time-of-flight detector are used for trigger and direction determination. The
Transition Radiation Detector and the Electromagnetic Calorimeter (Ecal) perform e/p separation and energy determination, and
a \v{C}erenkov counter measures charge and $\beta$.\\
The  principle of indirect search for dark matter is to look for non-standard signals in cosmic spectra, arising from
annihilations in the local halo \cite{bertone}. Such a deviation could be seen in the antimatter to matter ratio for charged
cosmic rays (positrons, antiprotons or antideuteron), for which AMS capabilities are very high \cite{alcaraz}. Positron
signal is of special interest since HEAT experiment has measured an excess which could be explained in terms of a Weakly
Interactive Massive Particle (WIMP) annihilation signal \cite{heat}. With a background rejection of order $10^{6}$, an
acceptance of $0.04\:m^{2}.sr$ and an energy resolution of $3\%$ in this range, AMS02 is awaited to confirm and precisely
measure the positron excess \cite{jon}. In the case of $\gamma$ rays, dark matter is expected to produce an enhancement of
diffuse emission from the halo and possibly point-like sources where its density is high. The Galactic center could be such a
source, in particular if the halo profile near the supermassive black hole is cusped. Another possibility is to observe a GeV
scale line emission, which would give a compelling evidence of the presence of dark matter since no known astrophysical
object could be able to produce it. Although originally designed for charged particles detection, AMS02 has high performance
in $\gamma$ rays detection. Two modes allow the spectrometer to perform a measurement whether the photon converts into a
$e^{+}/e^{-}$ pair in the upper detector or not. The Tracker is used in the first case and provide good angular resolution
($0.05^{o}$). A specific trigger is used in the other case \cite{moi}, for which the Ecal allows a 3\% energy resolution and
a $1^{o}$ angular resolution. These two complementary ways to detect high energy photons puts AMS02 in great place to perform
$\gamma$ ray astronomy, with high acceptance (of order $0.09\;m^{2}.sr$) and a large sky coverage \cite{loic}.

\section{Making micrOMEGAs an event generator}

\subsection{Architecture of the code}

The final goal of our package is to predict all indirect signals from SuSy dark matter \cite{micro_all}. The user may choose
to work in the general MSSM or with any  model constrained at the high scale (such as mSUGRA and AMSB), in that case the
first step is to compute the evolution of physical parameters such as masses and couplings from the unification scale to the
electroweak scale (EWS) at which cross sections in the halo have to be determined. The evolution from the high scale to EWS
is managed here by \verb"SuSpect" \cite{suspect} which solves the renormalization group equations \footnote{The results
presented here make use of SuSpect, but any other RGE code can be used}. All EWS parameters are given to \verb"micrOMEGAs"
which computes all cross sections. The former are discussed in next sections. At this point one has to link these final
states to observable particles, this task is devoted to \verb"PYTHIA" \cite{pythia}. It computes the partons
hadronization-fragmentation and the decays of unstable particles. For the $\gamma$ signal the integral of the density over a
field of view is computed. The charged particles have to be propagated through the Galaxy from the location of their
production to the Earth vicinity, then specific propagation codes are used. In the present note, $\bar{p}$ and $\bar{D}$
propagation are not presented, but they will be included in the final version.
\begin{figure}[H]
  \centering
  \includegraphics[width=17cm]{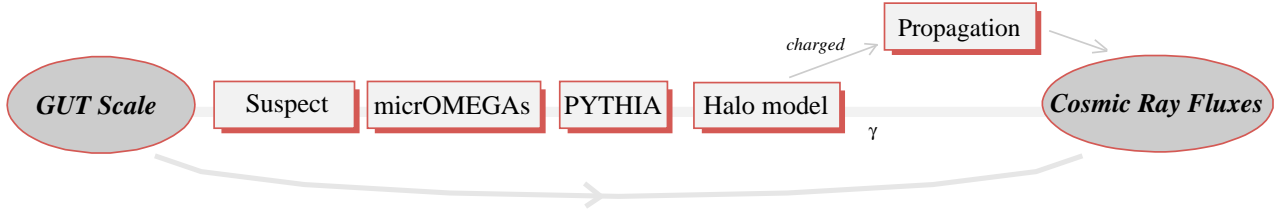}
  \caption{Principle of the code}
  \label{plan}
\end{figure}

\subsection{Fluxes from neutralino annihilations}
Dark matter signal in cosmic radiations can be seen either in the matter to antimatter ratio and in $\gamma$ rays. In this
note some results for $\gamma$ rays and $e^{+}$ are presented. Concerning the $\gamma$ rays from Galactic center, the flux is
given by the following expression:

\subsubsection*{\hspace{1cm}$\bullet$ $\gamma$ rays}
\begin{equation}
\Phi_{\gamma}^{SuSy}=\frac{dN_{\gamma}}{dSdEdtd\Omega}=\frac{1}{4\pi}\overbrace{\frac{dN_{\gamma}}{dE_{\gamma}}\frac{<\sigma_{ann}v>}{m_{\chi}^{2}}}^{Particle\;physics}\overbrace{\int_{l.o.s.}\frac{\rho_{\chi}^{2}(r(l_{\psi}))}{2}dl_{\psi}}^{Astrophysics}
\label{los}
\end{equation}
Two contributions are present in this equation, one purely astrophysical and the other arising from particle physics. The
right part is the density integral along a line of sight, this depends on the modelling of the dark halo. In the other part,
from particle physics, the fraction $<\sigma_{ann}v>/m_{\chi}^{2}$ contains the total annihilation cross section and the
neutralino mass, both provided by \verb"micrOMEGAs". Here the annihilation cross section has to be computed at the
temperature of the halo, whereas for the relic density estimation it was computed at $T_{freeze-out}$.
$dN_{\gamma}/dE_{\gamma}$ is the number of $\gamma$ per unit energy for one annihilation and is computed with \verb"PYTHIA"
(here version 6.123). Each final state giving different $\gamma$ spectra, this is in fact a sum over all final states :

\begin{equation}
\frac{dN_{\gamma}}{dE_{\gamma}}=\sum_{f.s.} \left[ B_{f.s.} \frac{dN_{\gamma}^{f.s.}}{dE_{\gamma}} \right ]
\end{equation}

As $\chi$ is electrically neutral, it does not couple to photons, therefore there is no tree level direct $\gamma$
production. Nevertheless, $\gamma$ rays arise from the hadronization of the final states partons. A typical example is the
decay of $\pi^{0}$'s appearing in the hadronization of a quark/antiquark pair. These processes lead to a continuous $\gamma$
spectrum. At the 1 loop level, $\gamma$ lines can be produced via processes such as $\chi\chi\rightarrow\gamma\gamma,\gamma
Z^{0}$. The $\gamma\gamma$ and $\gamma Z^{0}$ processes are implemented in our code \cite{fawzi}, and give photons at the
following discrete energies:

\begin{equation}
  \gamma\gamma\::\:E_{\gamma}=m_{\chi}\:(N_{\gamma}=2)\:\:\:;\:\:\: \gamma Z^{0}\::\:E_{\gamma}=m_{\chi}\left [
1- \left ( \frac{m_{Z^{0}}}{2m_{\chi}} \right )^{2} \right ]\:(N_{\gamma}=1)
\end{equation}

These processes are of special interest since no known astrophysical source could lead to GeV to TeV $\gamma$ lines. The
observation of such lines would provide a smoking gun signature for the observation of dark matter annihilations in the Milky
Way. In this note I present an application of the upcoming tool \verb"SloopS" developed by Boudjema \it{et al. }\rm that will
compute all of the MSSM processes at the 1 loop level. This package will be provided with \verb"micrOMEGAs".

\subsubsection*{\hspace{1cm}$\bullet$ Positrons}

For the positron signal, the flux is a solution of the propagation equation :
\begin{equation}
\frac{\partial}{\partial t} \frac{dn_{e^{+}}}{dE_{e^{+}}}= \vec{\nabla}. \left[
K(E_{e^{+}},\vec{x})\vec{\nabla}\frac{dn_{e^{+}}}{dE_{e^{+}}} \right ] + \frac{\partial}{\partial E_{e^{+}}} \left[
b(E_{e^{+}},\vec{x})\frac{dn_{e^{+}}}{dE_{e^{+}}} \right ] + Q(E_{e^{+}},\vec{x})
\end{equation}
with the source term :
\begin{equation}
Q(E_{e^{+}},\vec{x}) = \frac{\rho_{\chi}^{2}(\vec{x}))}{2} \frac{<\sigma v>}{m_{\chi}^{2}} \frac{dN_{e^{+}}}{dE_{e^{+}}}
\end{equation}

Here, \verb"micrOMEGAs" provides cross-sections and masses, \verb"PYTHIA" gives the differential $e^{+}$ spectrum and a new
code is used to solve the propagation equation, developed by Salati $et\: al.$ \cite{pierre_posit}. The different parts from
particle physics and astrophysics are detailed in the following sections.

\subsubsection{Cross sections and final states}
\begin{figure}[H]
  \centering
  \includegraphics[width=8cm]{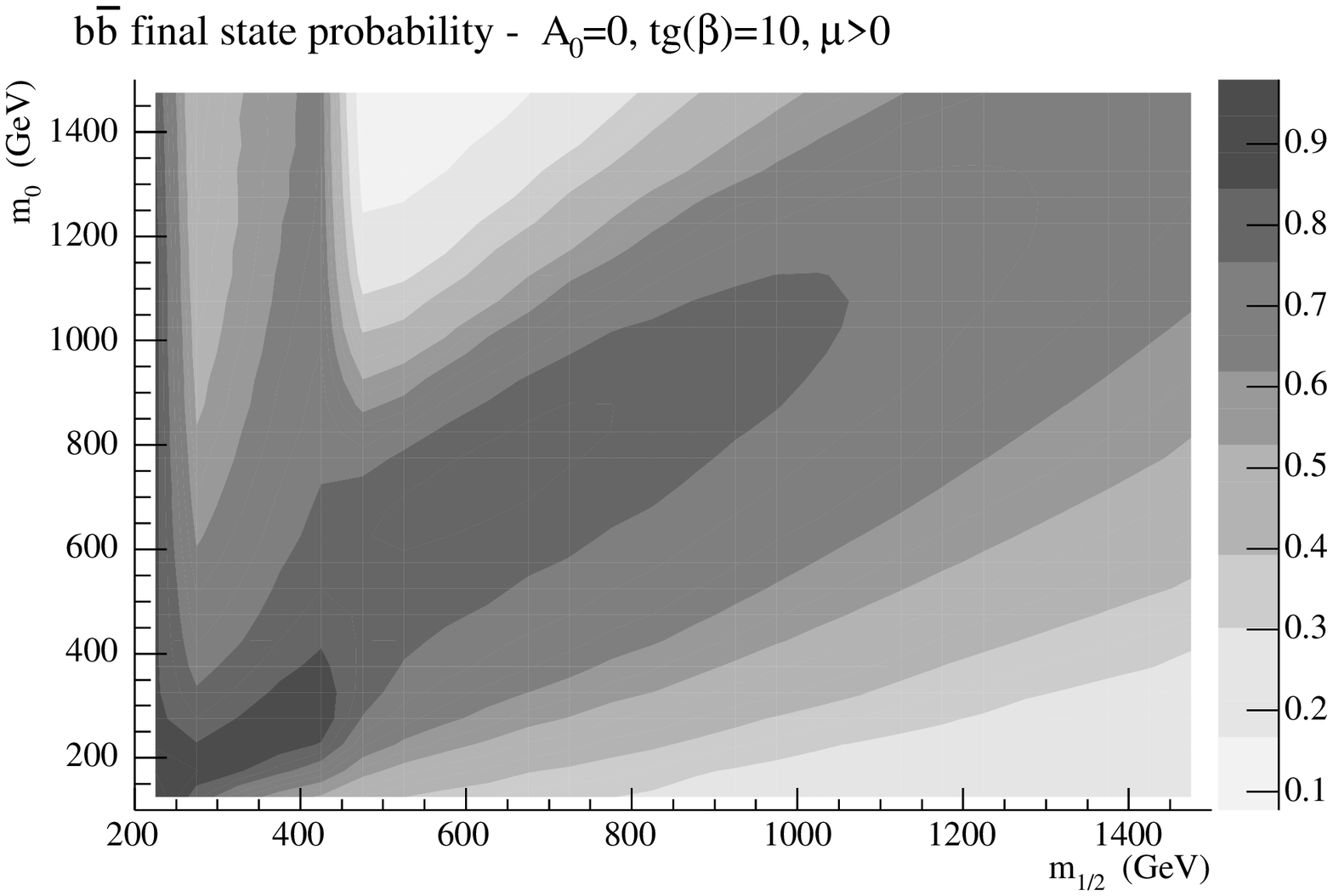}
  \includegraphics[width=8cm]{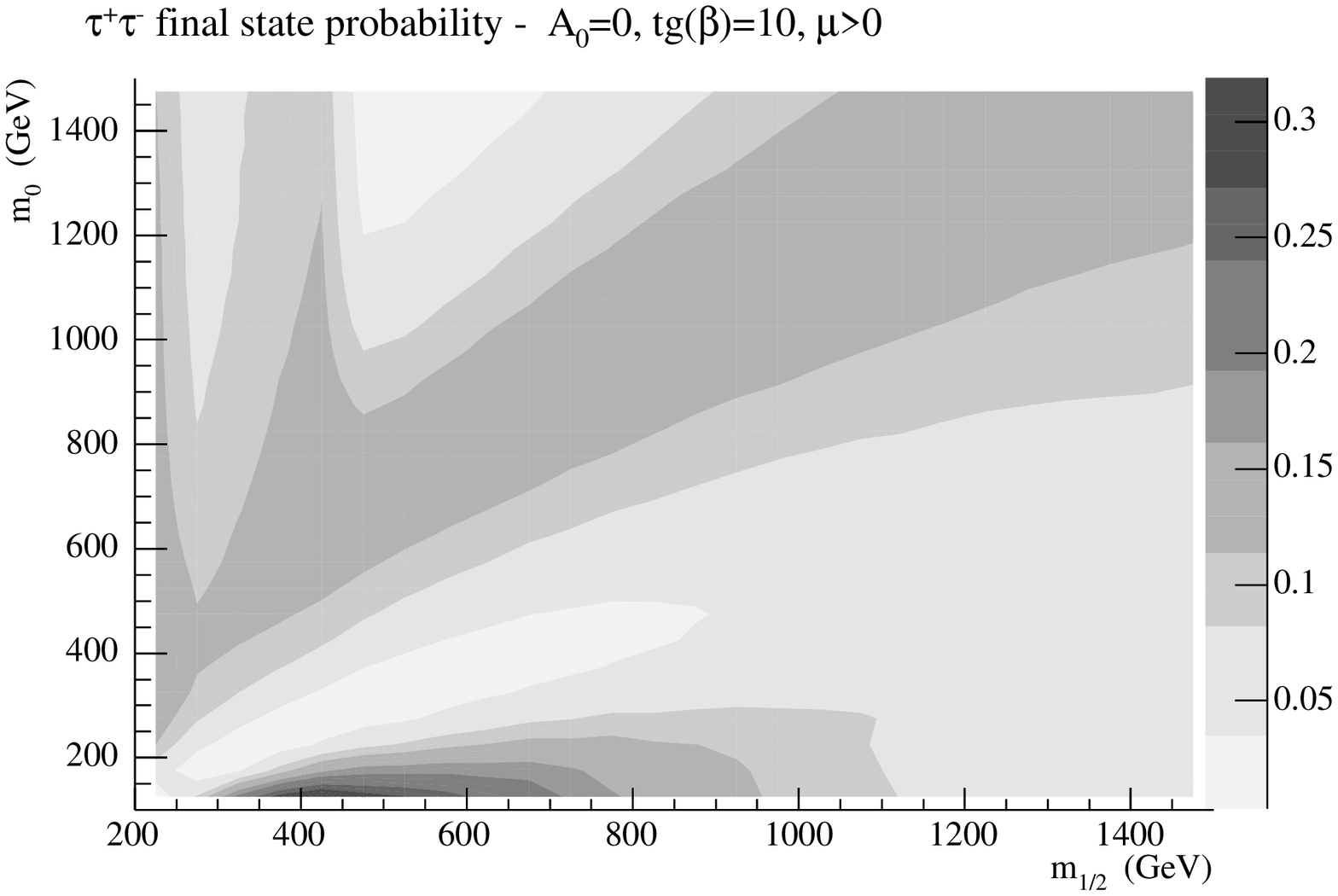}
  \includegraphics[width=8cm]{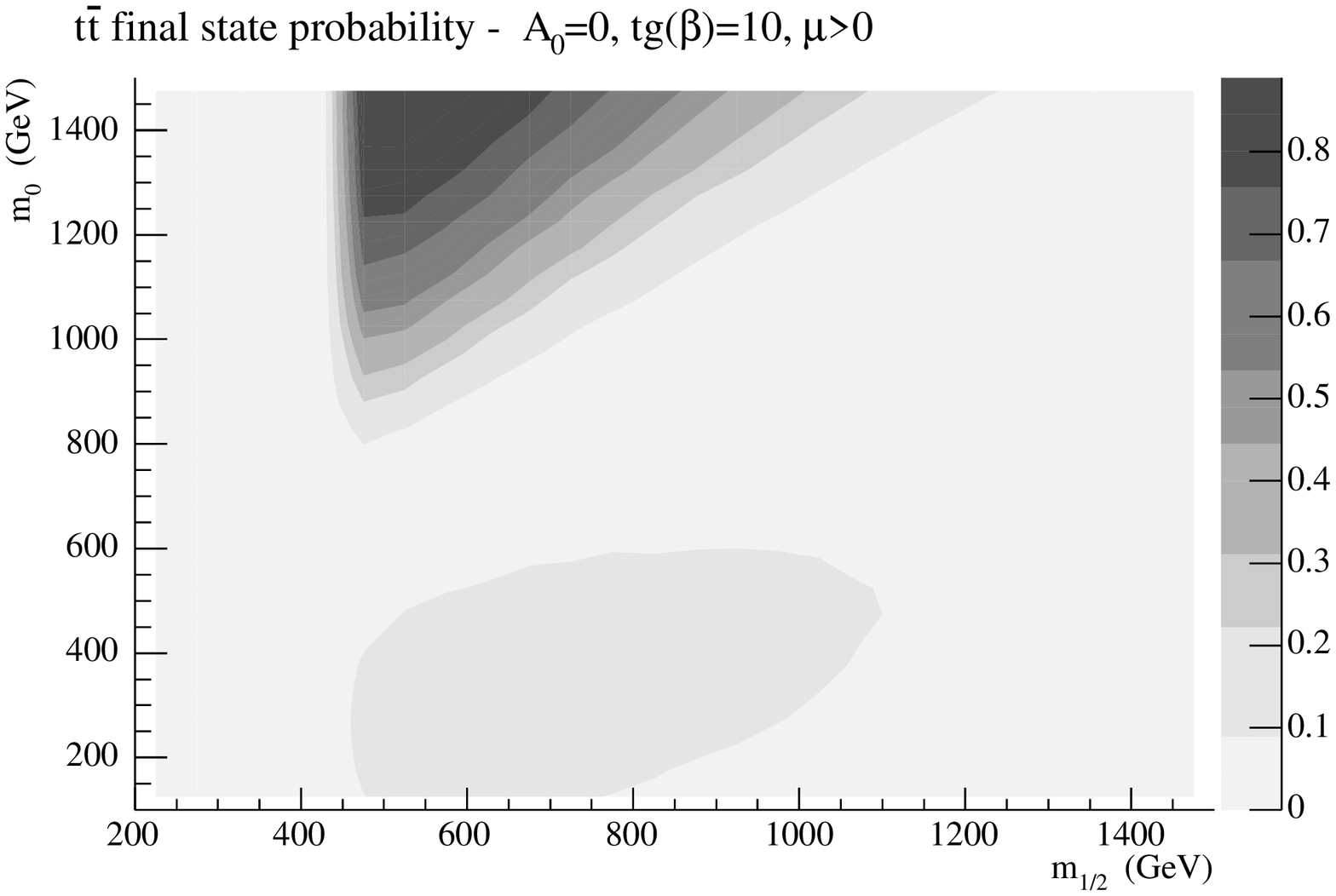}
  \includegraphics[width=8cm]{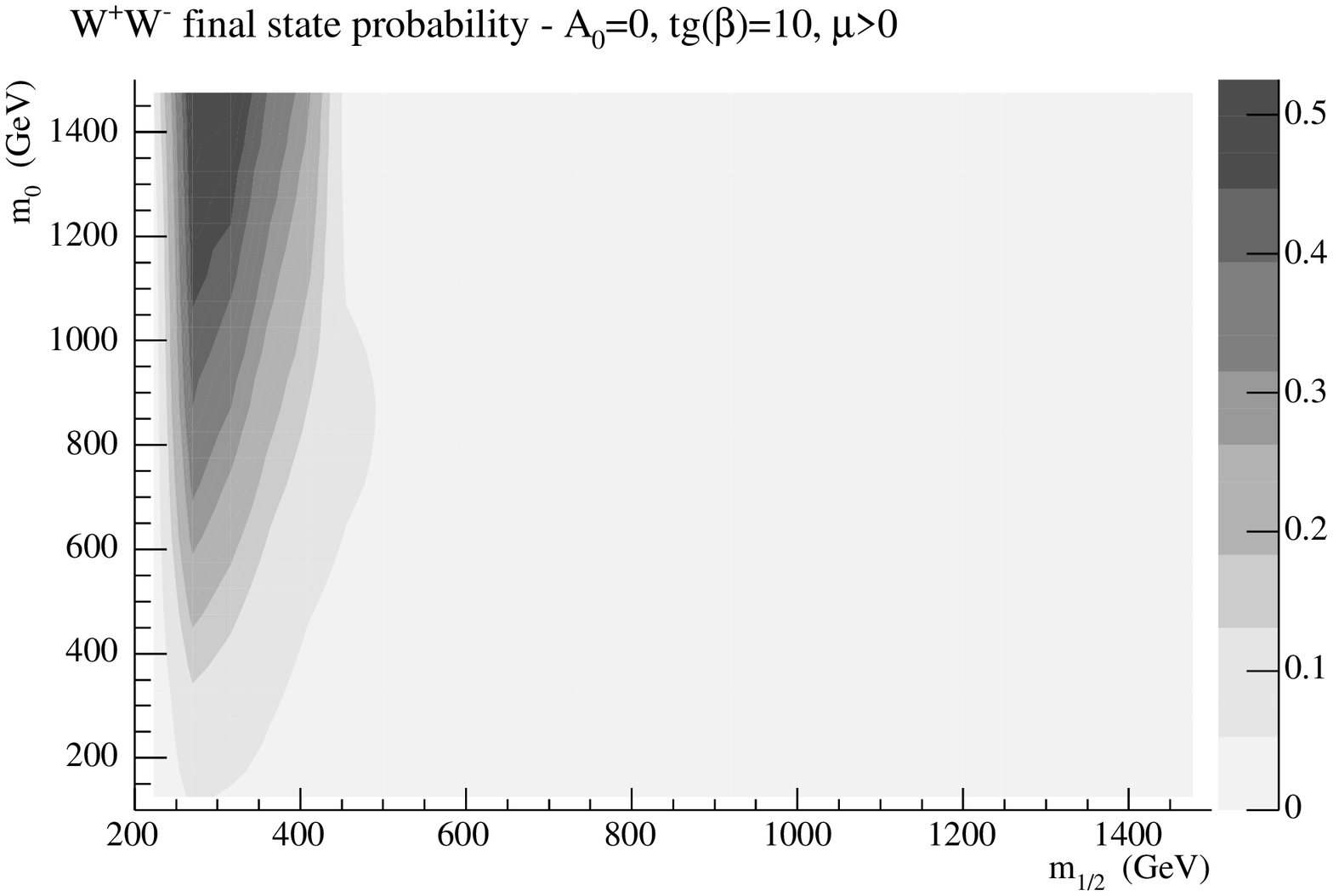}
  \caption{Final state occurrences with $A_{0}=0$, $\mu>0$ and $tg(\beta)=10$ in mSUGRA}
    \label{br10}
\end{figure}

There are 26 possible 2-body tree-level final states for neutralinos self annihilation. These are fermion/antifermion pairs
or allowed combinations of gauge and/or Higgs bosons. Each final state will give its own particles yields, with its specific
spectral features as we will see in section 4. Therefore it is important to know for a given model or set of parameters which
final states will be favored. In the mSUGRA framework, Figure \ref{br10} shows the occurrence of some final states
($b\bar{b}$, $t\bar{t}$, $\tau^{+}\tau^{-}$ and $W^{+}W^{-}$) in the $m_{0}$-$m_{1/2}$ plane with all other parameters fixed.

It appears from Figure \ref{br10} that a typical mSUGRA case is the one in which annihilation into a $b\bar{b}$ pair is
dominant, followed by the annihilation into $\tau$'s. Some regions of the parameter space lead to an annihilation into $top$
quarks, with a threshold effect, or into gauge bosons such as $W$'s. However, by comparing this figure to Figure \ref{cosmo},
one can see that the annihilation into $W^{+}W^{-}$ is not favored by cosmology in the case of mSUGRA. However it will be
shown in last section that other supersymmetry breaking scenarios can lead to this final state.

\subsubsection{Dark halo modelling}

\begin{figure}[H]
  \centering
  \includegraphics[width=9.5cm]{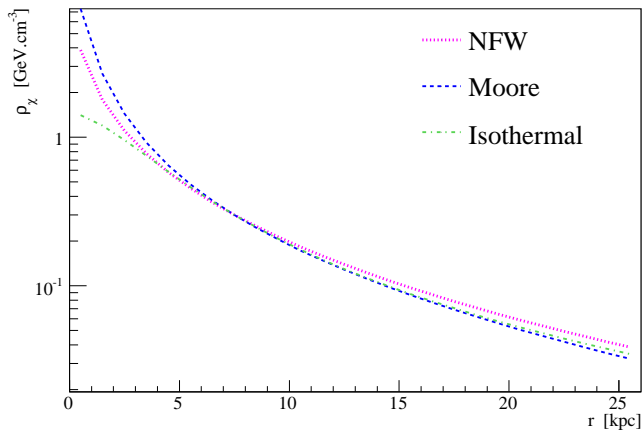}
  \caption{Three examples of halo profiles}
    \label{prof}
\end{figure}

The exact local dark matter density is not known and one has to parameterize the halo density profile in order to match the
observed gravitational effects. The local dark matter density is in the range $\rho_{\odot}=0.2-0.8\;GeV.cm^{-3}$ and that
does not give information about the density far from the center or near the center. Therefore one has to assume a halo
profile that gives the proper local density. The following parametrization is implemented in the code, it allows to describe
number of halo profiles (motivated by numerical simulations).
\begin{equation}
\rho_{\tiny CDM}(r)=\rho_{\odot} \left [ \frac{r_{\odot}}{r} \right ] ^{\gamma}
\left [ \frac{1+(r_{\odot}/a)^{\alpha}}{1+(r/a)^{\alpha}} \right ]
^{\frac{\beta-\gamma}{\alpha}}
\label{halo}
\end{equation}
In addition to this, any kind of central shape can be specified by hand, this is of special importance if one wishes to test
different accretion models onto the central supermassive black hole. Figures \ref{param} and \ref{prof} give the most popular
halo profiles parameters and shapes, in the case of Navarro-Frenk-White (NFW), Moore and isothermal sphere (see
\cite{bertone} and references therein).

\begin{figure}[H]
  \centering
    \begin{tabular}{c|c|c|c|c}
    Halo model & $\alpha$ & $\beta$ & $\gamma$ & a (kpc) \\
    \hline
    Isothermal with core & 2 & 2 & 0 & 4 \\
    \hline
    NFW & 1 & 3 & 1 & 20 \\
    \hline
    Moore  & 1.5 & 3 & 1.5 & 28 \\
    \end{tabular}
    \caption{Halo parameters for three profile types}
    \label{param}
\end{figure}

\begin{figure}[H]
  \centering
  \includegraphics[width=9.5cm]{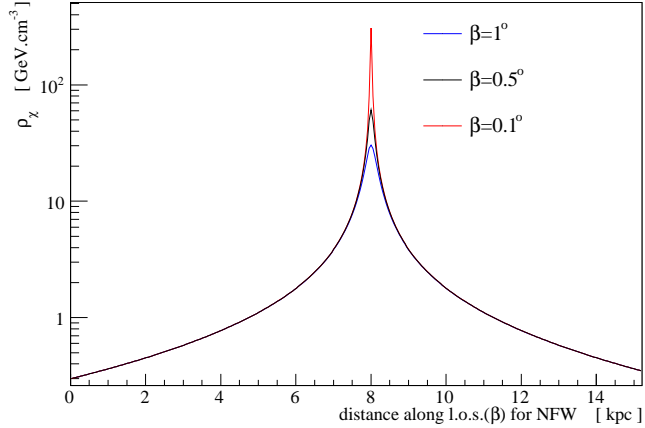}
  \caption{Dark matter density along lines of sight with different angles toward the Galactic center}
\label{losplot}
\end{figure}

This parametrization allows to predict diffuse fluxes from the halo as well as fluxes from the Galactic center.\\
In order to make a prediction for a $\gamma$ flux from the Galactic center, one has to integrate the signal inside some solid
angle around the direction of the Galactic center. To do so, the code sums all lines of sight contributions such as the ones
shown in Figure \ref{losplot}.

As an example, the table of Figure \ref{astro} show the normalized value of the astrophysical factor $<J>$, defined as the
mean value of J over a given $\Delta \Omega$ solid angle. Here $<J>$ is :
\begin{equation}
<J>_{\Delta \Omega}=\frac{1}{r_{\odot}\rho_{\odot}^{2}}\frac{1}{\Delta \Omega}\int d \Omega \int_{0}^{\infty}
\rho_{\chi}^{2}(r(l_{\psi}))dl_{\psi}
\end{equation}

\begin{figure}[H]
  \centering
  \begin{tabular}{c|c|c}
    Solid angle & $\Delta \Omega = 10^{-3}sr$ & $\Delta \Omega = 10^{-5}sr$ \\
    \hline
    Isothermal $<J>$&$2.621\;10^{1}$ &$2.624\;10^{1}$ \\
    \hline
    NFW $<J>$&$1.291\;10^{3}$ &$1.346\;10^{4}$ \\
    \hline
    Moore $<J>$&$1.947\;10^{5}$ &$1.562\;10^{7}$ \\
    \end{tabular}
    \caption{Astrophysical terms calculation with the package}
    \label{astro}
 \end{figure}

These figures are obtained with our package, they are in perfect agreement with previous calculations \cite{bertone}. These
figures show that in the case of this observation, a better angular resolution lead to a better observation only in the case
of a cuspy halo profile. Indeed for a halo with a flat dark matter distribution in its center, the mean value of the
astrophysical factor does not vary when the solid angle is changed.

\section{First example : $\gamma$ rays from Galacic center in mSUGRA}

I wish here to illustrate the use of our package for computing $\gamma$ spectra from the Galactic center. The astrophysical
part used here is a NFW parametrization (the distance to the Galactic center is $r_{\odot}=8\:kpc$  and the local dark matter
density is taken at $\rho_{\odot}=0.3\: GeV.cm^{-3}$). The annihilation rates are integrated inside a $1^{o}$ opening angle
cone around the Galactic Center. Then the $\gamma$ ray flux is given by the following formula :
\begin{equation}
  \Phi_{\gamma}^{Susy}(\Delta \Omega)\; \Delta \Omega=\frac{dN_{\gamma}}{dSdEdt}=\frac{1}{4\pi}\frac{dN_{\gamma}}{dE_{\gamma}}\frac{<\sigma_{ann}v>}{m_{\chi}^{2}}\int_{\Delta \Omega}{d\Omega}\int_{l.o.s.}{\frac{\rho_{\chi}^{2}(l)}{2}dl}
  \label{losint}
\end{equation}
As examples, I use two sets of mSUGRA with parameters detailed below. Both points match relic density and accelerator
constraints. The first one is used by W. de Boer $et\:al.$ \cite{deboer} and has the advantage of giving quite a high signal,
with a total cross section of order $10^{-26}\:cm^{3}s^{-1}$. The second point gives less flux but a visible line (G'
benchmark point \cite{ellis}). Figure \ref{tab1} gives some characteristics of the chosen sets of parameters.
\begin{figure}[H]
  \centering
  \begin{tabular}{c||c}
    Set 1 & Set 2\\
    \hline
    $m_{0}=m_{1/2}=A_{0}=500\:GeV$ & $m_{0}=113\:GeV$, $m_{1/2}=375\:GeV$, $A_{0}=0$\\
    $tg(\beta)=50$, $\mu>0$ & $tg(\beta)=20$, $\mu>0$\\
    \hline
    $\Omega_{\chi}h^{2}=0.098$ & $\Omega_{\chi}h^{2}=0.128$\\
    $m_{\chi}=206.9\:GeV$ & $m_{\chi}=151.5\:GeV$\\
    $<\sigma_{tot}v>=1.81\;.10^{-26}\:cm^{3}s^{-1}$ & $<\sigma_{tot}v>=6.96\;.10^{-28}\:cm^{3}s^{-1}$\\
    $<\sigma(b\bar{b})v>=0.87\;\times<\sigma_{tot}v>$ & $<\sigma(b\bar{b})>=0.58\;\times<\sigma_{tot}v>$\\
    $<\sigma(\tau^{+}\tau^{-})v>=0.13\;\times<\sigma_{tot}v>$ & $<\sigma(\tau^{+}\tau^{-})>=0.39\;\times<\sigma_{tot}v>$\\
    $<\sigma(\gamma\gamma)v>=2.10^{-5}\;\times<\sigma_{tot}v>$ & $<\sigma(\gamma\gamma)>=0.01\;\times<\sigma_{tot}v>$\\
    $<\sigma(\gamma Z^{0})v>=4.10^{-6}\;\times<\sigma_{tot}v>$ & $<\sigma(\gamma Z^{0})>=0.001\;\times<\sigma_{tot}v>$\\
  \end{tabular}
  \caption{Main features of the two sets of mSUGRA parameters considered in this part}
    \label{tab1}
\end{figure}
All the final states cross sections being determined, the $\gamma$ spectra for each one of them has to be computed. This is
done by using \verb"micrOMEGAs" final states as an input to \verb"PYTHIA", which hadronizes the partons, simulates the jets
fragmentation and the decay of all unstable particles that may appear in the hadronization process. Different spectra are
obtained for each final states. The final spectra results in weighting these ones with the computed cross sections. The user
has different ways to obtain the final signal:
\begin{itemize}
  \item{
    Monte Carlo generator: The spectrum is produced event by event with a final state random choice according to their
    probability.
  }
  \item{
    Weighted channels: Channels are produced separately and then weighted as regard to their cross section.
  }
  \item{
    Use of tables : A catalog of spectra is available, those being interpolated to
    match the specified parameters. Although not very accurate, this method is much faster and allows a fast
    rough determination of the spectra, which is useful for scans of the parameter space.
  }
\end{itemize}
In this paper the second method is used and all channels with probability greater than 0.1\% plus $\gamma\gamma$ and $\gamma
Z^{0}$ have been considered in the signal prediction.
\newpage
\subsubsection*{Set 1}
For the first set of parameters, the following spectrum for the $\gamma$ rays arising from Galactic center is obtained. Here
the flux is high and the $\gamma$ lines are undistinguishable from the continuous hadronization spectrum.
\begin{figure}[H]
\centering
\includegraphics[width=9cm]{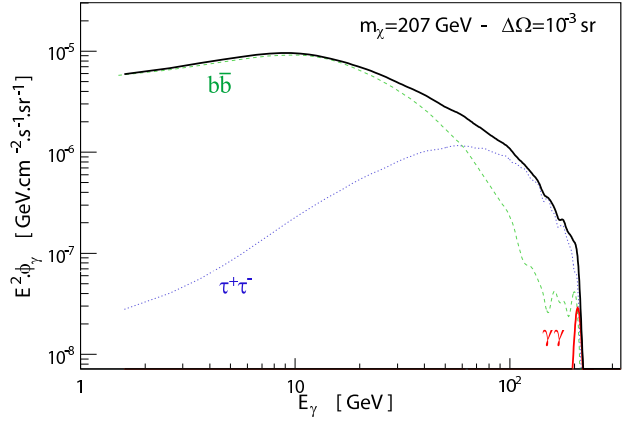}
\caption{$\gamma$ flux from Galactic center for set 1}
\label{spec1}\end{figure}

\subsubsection*{Set 2}

\begin{figure}[H]
\centering
\includegraphics[width=9cm]{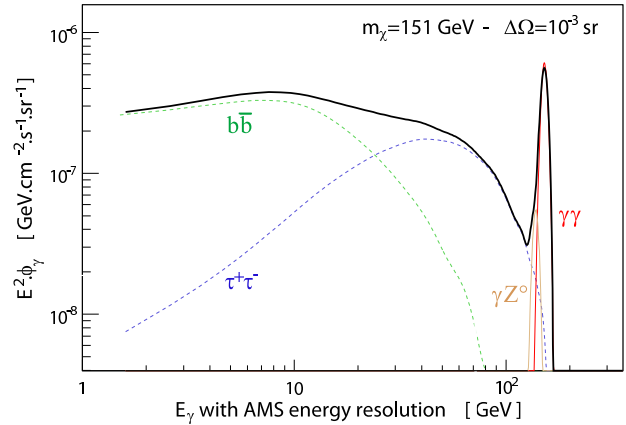}
\caption{$\gamma$ flux from Galactic center for set 2} \label{spec2}
\end{figure}

For the second set of parameters, Figure \ref{spec2} is obtained. As expected the flux is quite low but one can see very
clearly the lines induced by $\chi\chi\rightarrow\gamma\gamma$ and $\chi\chi\rightarrow\gamma Z^{0}$.

\begin{figure}[H]
\centering
\includegraphics[width=9cm]{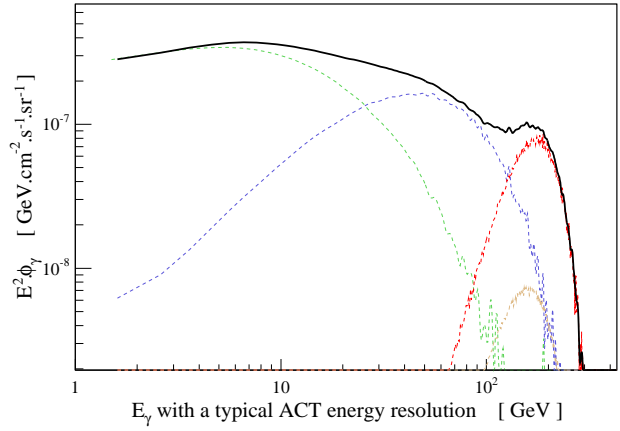}
\caption{Same as previous Figure \ref{spec2} with an ACT energy resolution} \label{act}\end{figure}

Of course, those two spectra have to be added to the background and compared to AMS02 acceptance. According to the background
expectation, it will be less obvious to extract the signal in the latter case (in next section a estimation of AMS02
sensitivity to this signal is presented). One can notice that the specific spectral features can only be observed with a very
good energy resolution. In order to illustrate this, one can plot the same spectrum with an typical Atmospheric \v{C}erenkov
Telescope (ACT) energy resolution, as it is done in Figure \ref{act} (a $\sim$30\% energy resolution is assumed).

In the case of an ACT, the flux suppression at $m_{\chi}$ could be seen, but the $\gamma$ line itself is too much smeared by
the energy resolution. Only a space mission like AMS02 would be able to do so.

\subsubsection*{Estimation of theses fluxes observability with AMS02}

\begin{figure}[H]
\centering
\includegraphics[width=9cm]{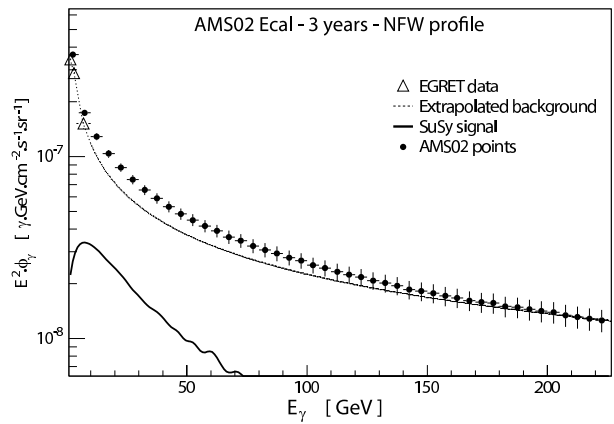}
\caption{AMS02 measurement of the flux from Galactic center for set 1} \label{set1_bkg}
\end{figure}

As it was said before, the first set of parameters provides quite a high signal. In order to have a rough estimate of the
observability of this flux with AMS02, an average acceptance of $0.09\:m^{2}.sr$ is assumed, and the signal is integrated
inside a solid angle of $10^{-3}\:sr$ corresponding to the $1^{o}$ angular resolution of the Ecal. For 3 years of data
taking, the measurement of the  $\gamma$ flux in case of set 1 would lead to the points of Figure \ref{set1_bkg}. Here the
expected background is a power law of spectral index $2.72$ extrapolated out of the EGRET measurement of $\gamma$ rays from
the Galactic center region \cite{egret}.

\begin{figure}[H]
\centering
\includegraphics[width=9cm]{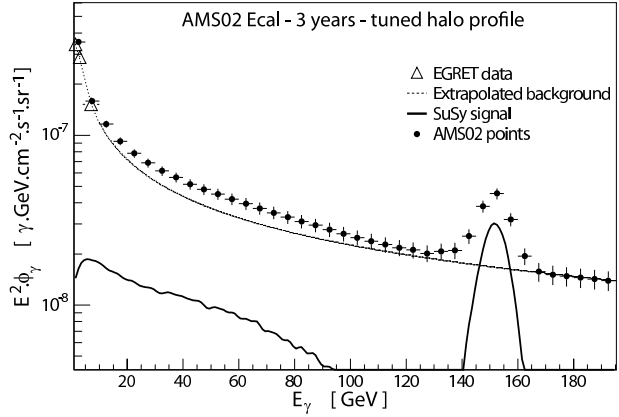}
\caption{AMS02 measurement of the flux from Galactic center for set 2, with a specific halo profile} \label{set2_bkg}
\end{figure}

In the second scenario, the signal shown on Figure \ref{spec2} gives a signal that leads to an excess at the level of less
than $1\sigma$ over the background, even in 10 years of data taking. It would be possible to observe this line if the dark
matter halo were more cusped than a typical NFW profile. In Figure \ref{set2_bkg}, the observation of the line by AMS02 in 3
years is presented under the more optimistic assumption of a central cusp of index $\gamma=1.25$.

\section{Second example : positron signal in AMSB}

In the AMSB scenario, the neutralino is a pure wino, and the annihilation cross section is always higher than in the mSUGRA
case. This has two main implications :
\begin{itemize}
\item{
    The thermal relic density is very small so that one has to assume non thermal $\chi$ production.
    Such a non-thermal relic density can be caused by non-standard cosmology \cite{pierre_quin} or decays of gravitinos produced
    at the end of inflation \cite{randall2}.}
\item{
  The $\chi\chi\rightarrow W^{+}W^{-}$ channel is always dominant as soon as it is kinematically allowed.
}
\end{itemize}

The annihilation into W bosons gives harder $e^{+}$ signal than in the typical $b\bar{b}$ case of mSUGRA. That is why this
framework has been chosen to illustrate the generation of positron signal. In the example here, the SPS9 benchmark point
\cite{snowmass}is used, for which :

\begin{figure}[H]
  \centering
  \begin{tabular}{|c|}
    \hline
    SPS 9 \\
    \hline
    $m_{0}=450\:GeV,\:\:\:m_{3/2}=60\:TeV,\:\:\:tg(\beta)=10,\:\:\:\mu>0$\\
    \hline
    $(\Omega_{\chi}h^{2})_{thermal}=0.0018$\\
    $m_{\chi}=197.6\:GeV$\\
    $<\sigma_{tot}v>=1.93\;.10^{-24}\:cm^{3}s^{-1}$\\
    $<\sigma(W^{+}W^{-})v>=0.991\;\times<\sigma_{tot}v>$\\
    $<\sigma(\gamma\gamma)v>=0.0021\;\times<\sigma_{tot}v>$\\
    $<\sigma(\gamma Z^{0})v>=0.0065\;\times<\sigma_{tot}v>$\\
    \hline
  \end{tabular}
  \label{tableau}
  \caption{Main features of the SPS9 set of parameters}
\end{figure}

\begin{figure}[H]
\centering
\includegraphics[width=8.5cm]{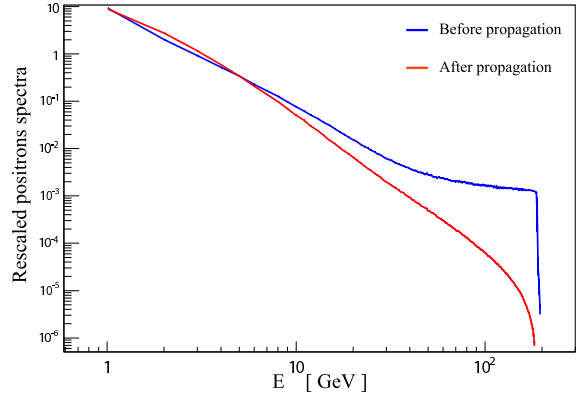}
\caption{Positron source term and flux after propagation (see text)} \label{posit1}
\end{figure}

For the flux calculation, the same halo model as above is used. Figure \ref{posit1} shows the $e^{+}$ before propagation and
after propagation. The code that was used here is currently under development \cite{pierre_posit} and its final version will
be implemented in \verb"micrOMEGAs". It makes use of a new positron propagator for the diffusion in the galaxy and its
spatial resolution should not be limited, in order to allow the study of small dark matter structures. However for the signal
shown here, the $e^{+}$ source term is integrated over the whole galactic halo, taken to be smooth.

\begin{figure}[H]
\centering
\includegraphics[width=9cm]{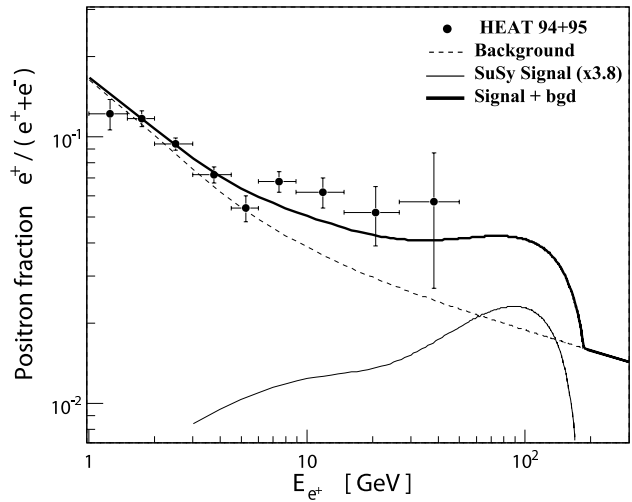}
\caption{Fit to the HEAT data in AMSB} \label{posit2}
\end{figure}

Figure \ref{posit1} shows the superimposition of the $e^{+}$ source term (blue curve) and the flux after propagation. Here
the red curve has been re-scaled in order to see the spectrum distortion due to propagation. The scale on the left
corresponds to the source term in units $positrons.GeV^{-1}$. The scaling factor used here to draw the flux curve is of order
$10^{6}\;cm^{2}.s.sr$. Dark matter substructures (clumps) are awaited to produce an enhancement of the signal
\cite{pierre_posit}. Because of the high annihilation cross-section and the specific high energy features of the W-induced
positrons, a boost factor of only order unity is required to fit the HEAT positrons excess \cite{heat}. Figure \ref{posit2}
shows the SuSy signal in term of positron fraction, compared to the HEAT excess. The expected background is taken from
\cite{posit}.

In Figure \ref{posit2} the supersymmetric signal is multiplied by a factor of 3.8, this figure is to be compared to the
typical mSUGRA case, in which the boost factor can vary from 10 to over 1000. An important feature of the AMSB scenario is
the charginos/neutralinos mass degeneracy, which implies a very low sensitivity for hadron colliders to this model. Therefore
if AMSB is the proper description of high energy physics, AMS02 would be in a very good position to observe SuSy.

\section{Other charged channels : antiprotons and antideuterons}

\begin{figure}[H]
\centering
\includegraphics[width=7.5cm]{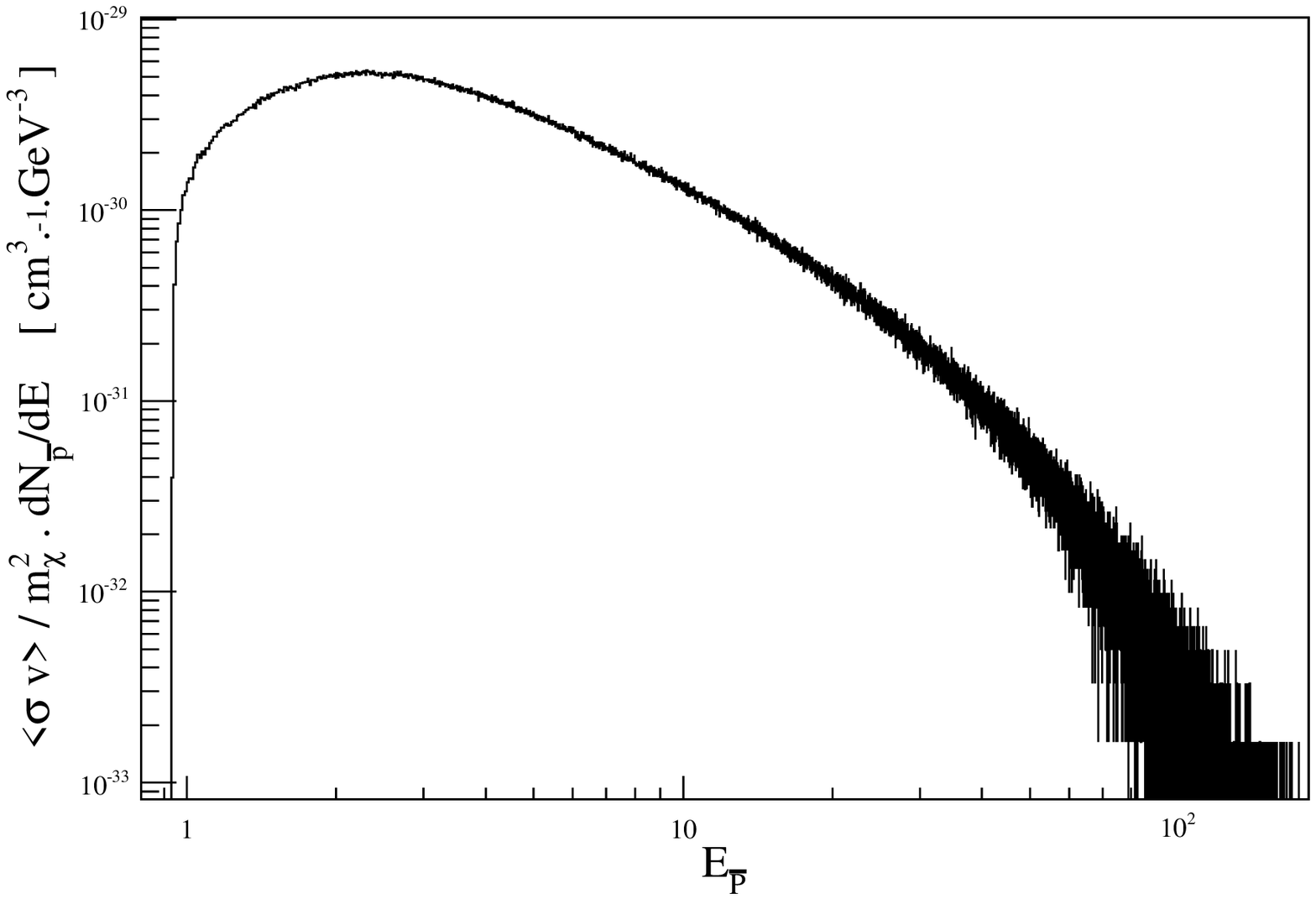}
\includegraphics[width=7.5cm]{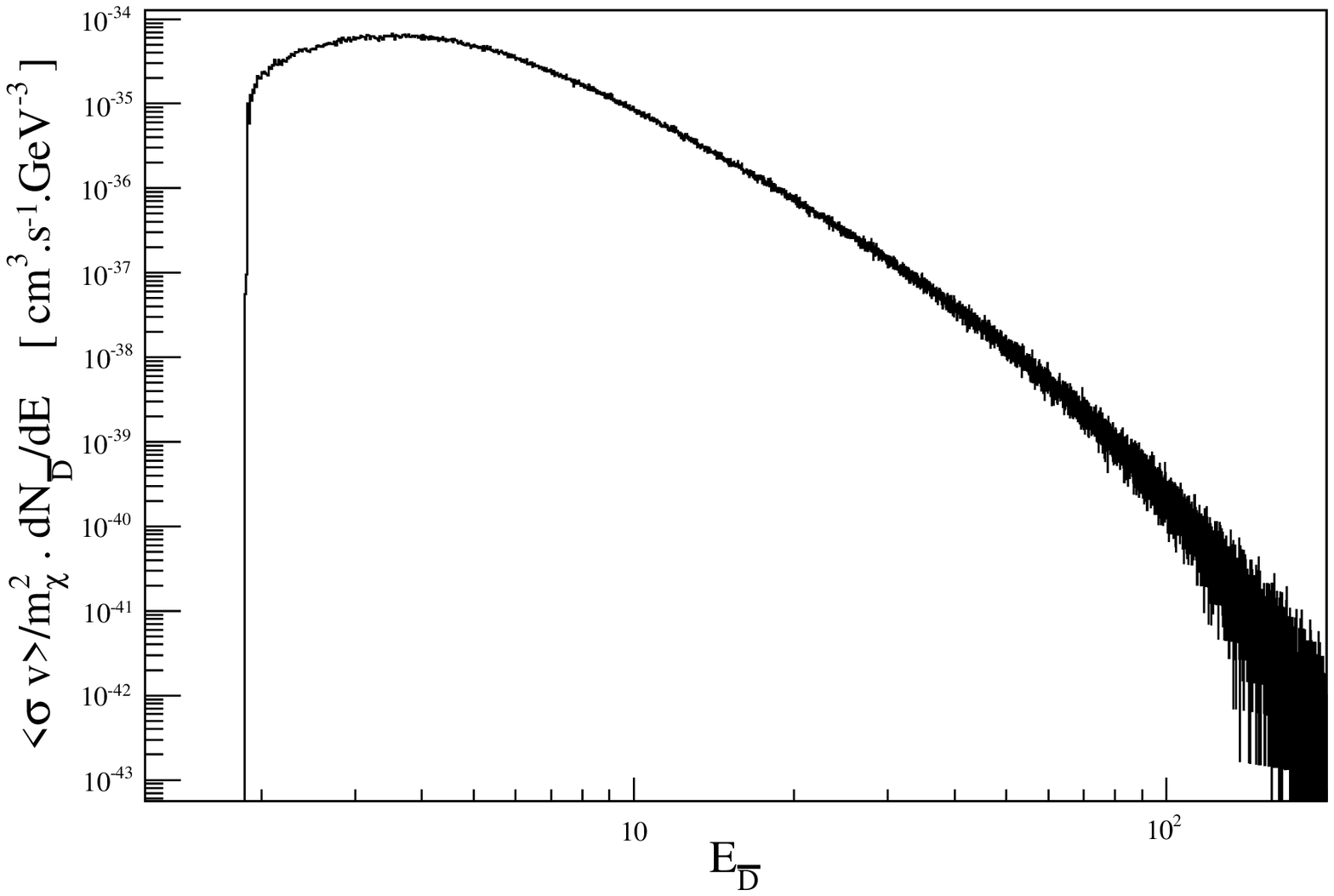}
\caption{$\bar{p}$ and $\bar{D}$ source spectra} \label{sources}
\end{figure}

For the moment, the propagation of $\bar{p}$ and $\bar{D}$ is not implemented in the code. However, the injection spectra
that stands for the source terms in the propagation equations are available. Figure \ref{sources} shows the source terms for
these two channels.

The $\bar{D}$ spectrum is computed according to the coalescence model, see ref \cite{dbar}. The adjunction of these channels
in the code is currently under way.

\section{Conclusions and outlook}

In this note I gave some examples of SuSy signals which could be seen by AMS02. These illustrations show the potential of the
package we are developing. Its possibilities will be even greater than what is shown here, it allows to work in any SuSy
model and with any halo profile. In the final version, substructures and extragalactic sources studies will also be possible.
Some comparisons with the only existing similar tool DarkSUSY \cite{darksusy} are being performed and we are currently
working on the implementation of the $\bar{p}$ and $\bar{D}$ propagation. In the near future, the nature of dark matter could
be unveiled by the spectral features of the annihilation signals in the Milky Way. Once published, this package could be a
powerful tool for data analysis of all experiments performing indirect searches of dark matter.

\section*{Acknowledgements}

I would like to thank all the members of LAPP and LAPTH involved in this project for this fruitful and pleasant
collaboration, as well as all the French working groups GDR SUSY and PCHE members.

\end{document}